# Quality Control and Due Diligence in Project Management: Getting Decisions Right by Taking the Outside View

By

Bent Flyvbjerg

Professor, Chair, Director

BT Centre for Major Programme Management

Saïd Business School

University of Oxford

bent.flyvbjerg@sbs.ox.ac.uk

Version 5.2

# Quality Control and Due Diligence in Project Management: Getting Decisions Right by Taking the Outside View[1]

## Abstract

This paper explores how theories of the planning fallacy and the outside view may be used to conduct quality control and due diligence in project management. First, a much-neglected issue in project management is identified, namely that the front-end estimates of costs and benefits – used in the business cases, cost-benefit analyses, and social and environmental impact assessments that typically support decisions on projects – are typically significantly different from actual ex post costs and benefits, and are therefore poor predictors of the actual value and viability of projects. Second, it is discussed how Kahneman and Tversky's theories of the planning fallacy and the outside view may help explain and remedy this situation through quality control of decisions. Third, it is described what quality control and due diligence are in the context of project management, and an eight-step procedure is outlined for due diligence based on the outside view. Fourth, the procedure is tested on a real-life, multibillion-dollar project, organized as a public-private partnership. Finally, Akerlof and Shiller's recent discussion in economics of "firing the forecaster" is discussed together with its relevance to project management. In sum, the paper demonstrates the need, the theoretical basis, a practical methodology, and a real-life example for how to de-bias project management using quality control and due diligence based on the outside view.

*Keywords*: Quality control, due diligence, front-end management, planning fallacy, outside view, forecasting, cost-benefit analysis, ex-post analysis, Daniel Kahneman.

## "Third Wave" Research

The 2011 *Oxford Handbook of Project Management* identifies a "third wave" in project management research characterized by a positioning of the management of projects as a vital part of general management (Morris, Pinto, and Söderlund 2011:3). With projects taking center stage as delivery model for products and services in most organizations, project management can no longer be seen as a specialist subcategory of management brought to bear on special cases, but must instead be seen as a core business activity, vital to organizations as a whole. As part of this development from specific to general management, theory has developed from project-based theory to more general theory, substantially strengthening the academic base of project management, according to Söderlund (2011) and Turner, Pinto, and Bredillet (2011). The present paper is intended as a contribution to, first, the third wave's focus on general management and, second, fortifying the theoretical basis of project management by focusing on general theory and its importance to project management.

Winter et al. (2006); Williams, Samset, and Sunnevåg (2009); Edkins et al. (2012); and many others have identified high-quality front-end management as crucial to good project delivery. The front-end has been singled out as the perhaps most important stage in the overall project cycle in securing the success of projects, or avoiding failure. Within the front-end, estimating expected outcomes of projects and evaluating whether they justify investing in a given project is a key activity, and maybe the most consequential one, because ideally estimates inform the decision of whether projects will go ahead or not.[2] Furthermore, the quality of estimates will highly influence whether a project is deemed successful or not in meeting its objectives. Yet, as pointed out by Edkins et al. (2012:6) there is little about this in the research literature, nor about how the quality of estimation may be improved. It is the purpose of the present paper to help fill this gap in the literature.

Specifically, the paper brings Nobel prize-winning economic theory about the planning fallacy and the outside view, developed by Daniel Kahneman and Amos Tversky, to bear upon quality control and due diligence in the management of projects. Previous studies have documented a much-neglected topic in research and practice, namely that ex ante estimates of costs and benefits – used in the business cases, cost-benefit analyses, and social and environmental impact assessments that typically support decisions on projects – are often significantly different from actual ex post costs and benefits (Flyvbjerg 2009). As a result, such business cases, cost-benefit analyses, and social and environmental impact assessments are typically poor predictors of the actual value and viability of projects and cannot be trusted as a basis for informed decision making. The idea proposed in the present article is:

1. to develop quality control and due diligence to establish just how reliable, or unreliable, the business case, cost-benefit analysis, and social and environmental impact assessment are for a given project, and
2. to decide the implications of this for the project go-ahead.

Kahneman and Tversky did not develop their theories with these purposes in mind. The thesis is, however, that their theories will be highly effective in quality control and due diligence of planned projects and thus in improving decision making on projects. The existing literature on quality control, due diligence, and risk assessment in project management does not cover this particular approach to improved decision making (Breakwell 2007; De Mayer, Loch, and Pich 2002; Fischoff 2002; Perminova, Gustafsson, and Wikström 2008; Winch 2010).

Due diligence is generally an evaluation of the assets of a company, an investment, or a person. Here, due diligence is specifically understood as an evaluation of the costs and benefits deriving from investing in a given project, and especially whether the estimated costs and benefits for that project are likely to materialize. Due diligence is thus developed to be used as "quality control" of business cases, cost-benefit analyses, and the go-decision in projects. Data to test the thesis are drawn from a large database on estimated and actual costs and benefits in major projects and programs. Theory and methodology are developed at a level of generalization sufficiently high to allow their use, not only in business cases and cost-benefit analyses, but also in social and environmental impact assessments, which are typically also key in decision making on major projects, and which also suffer from the problem of predicted impacts often being very different from actual ones (Flyvbjerg, Bruzelius, and Rothengatter 2003:49-64).

## The Planning Fallacy and the Outside View

Flyvbjerg, Holm, and Buhl (2002, 2005) and Flyvbjerg, Bruzelius, and Rothengatter (2003) showed that forecasts of costs and benefits of major projects are generally highly inaccurate and biased. These findings have been verified by other similar research (Arena et al 2006; Dantata, Touran, and Schneck 2006; Flyvbjerg and Stewart 2012; Gardener and Moffat 2008; Merrow 2011; Moløkken and Jørgensen 2003; Scott, Levitt, and Orr 2011; Williams, Samset, and Sunnevåg 2009). Kahneman and Tversky (1979a, 1979b), in their Nobel-prize-winning work on decision making under uncertainty, argued that such inaccuracy is caused by a systematic fallacy in decision making causing people to underestimate the costs, completion times, and risks of planned actions, whereas people overestimate the benefits of the same actions. Following and expanding upon Buehler, Griffin, and Ross (1994), Lovallo and Kahneman (2003:58) would later call such common behavior the "planning fallacy."[3] Kahneman (1994) argued that this fallacy stems from actors taking an "*inside view*" focusing on the constituents of the specific planned action rather than on the outcomes of similar actions already completed. Kahneman also identified a cure to the fallacy, namely taking an "*outside view*" on planned actions, which consists in using experience from previous similar ventures already completed, including (a) the average outcome in sets of such ventures and (b) distributional information about outcomes. Distributional information is here understood as data on variation from the average outcome, for instance as expressed in common statistical measures such as standard deviation and variance. For instance, students asked how long they will take to write their thesis typically significantly underestimate the time, trying to understand and estimate the task inside-out. If they would simply collect data and calculate the average of how long it took 10-15 fellow students in last year's class in the same program they would arrive at a much more accurate estimate, according to Kahneman. Below we will see how doing so in a systematic fashion may be used for performing due diligence of cost and benefit estimates, and for uncovering just how biased and error-prone such estimates typically are.

The theoretical and methodological foundations of the outside view were first described by Kahneman and Tversky (1979b). As explained in Flyvbjerg (2006), thinking about the outside view was originally developed to compensate for the type of cognitive bias that Kahneman and Tversky had found in their work. This work shows that errors of judgment are often systematic and predictable rather than random, manifesting bias rather than confusion, and that any corrective prescription should reflect this. It also shows that many errors of judgment are shared by experts and laypeople alike. Finally the work demonstrates that errors remain compelling even when one is fully aware of their nature. Thus awareness of a perceptual or cognitive illusion does not by itself produce a more accurate perception of reality, according to Kahneman and Tversky (1979b:314). Awareness may, however, enable one to identify situations in which the normal faith in one's impressions must be suspended and in which judgment should be controlled by a more critical evaluation of the evidence. Here it is argued that due diligence is a method for such critical evaluation. Human judgment, including in developing business cases and forecasts for projects, is biased. Due diligence is a method for systematically and reliably assessing how large biases are and for de-biasing business cases and other forecasts.

With Kahneman and Tversky (1979b) it is argued that the prevalent tendency for humans to underweigh or ignore distributional information is perhaps the major source of error in forecasting and decision making. "The analysts should therefore make every effort to frame the forecasting problem so as to facilitate utilizing all the distributional information that is available," say Kahneman and Tversky (1979b:316). This may be considered the single most important piece of advice regarding how to increase accuracy in business cases and forecasting through improved methods (as opposed to through improved incentives, see Flyvbjerg 2007a). Using such distributional information from other ventures similar to that being forecasted is taking an outside view and it is the cure to the planning fallacy, according to Kahneman and Tversky. Due diligence as described below is a method for consistently taking an outside view on planned actions.

The contrast between inside and outside views has been confirmed by systematic research (Buehler, Griffin, and Ross, 1994; Gilovich, Griffin, and Kahneman, 2002). The research shows that when people are asked simple questions requiring them to take an outside view, their forecasts become significantly more accurate. For example, a group of students enrolling at a college were asked to rate their future academic performance relative to their peers in their major. On average, these students expected to perform better than 84 percent of their peers, which is logically impossible. The forecasts were biased by overconfidence. Another group of incoming students from the same major were asked about their entrance scores and their peers' scores before being asked about their expected performance. This simple diversion into relevant outside-view information, which both groups of subjects were aware of, reduced the second group's average expected performance ratings by 20 percent. That is still overconfident, but it is significantly more realistic than the forecast made by the first group (Lovallo and Kahneman 2003:61).

However, most individuals and organizations – including project managers, cost engineers, and risk departments – are inclined to adopt the inside view in planning new projects. This is the conventional and intuitive approach. The traditional way to think about a complex project is to focus on the project itself and its details, to bring to bear what one knows about it, paying special attention to its unique or unusual features, trying to predict the events that will influence its future. The thought of going out and gathering simple statistics about related projects seldom enters a planner's mind, according to Kahneman and Tversky, who use the term "planner" to denote any individual or organization contemplating future actions. For this reason, biases in decision making are persistent and de-biasing will not happen of its own accord. For de-biasing to take place, incentives and methods for quality control of projects and decisions must be in place, like those described below.

While understandable, planners' preference for the inside view over the outside view is unfortunate. When both forecasting methods are applied with equal skill, the outside view is much more likely to produce a realistic estimate, as shown by the simple experiments conducted by Kahneman and others. That is because the outside view bypasses cognitive and political biases such as optimism bias and strategic misrepresentation and cuts directly to outcomes (Flyvbjerg 2007a). In the outside view project planners and forecasters are not required to make scenarios, imagine events, or gauge their own and others' levels of ability and control, so they cannot get all these things wrong. Human bias is bypassed. Undoubtedly, the outside view, being based on historical precedent, may fail to predict extreme outcomes, that is, those that lie outside all historical precedents. But for most projects, the outside view will produce more accurate results. In contrast, a focus on inside details will lead to inaccuracy and poor decisions.

The comparative advantage of the outside view will be most pronounced for non-routine projects, understood as projects that managers and decision makers in a certain organization or locale have never attempted before – like building greenfield infrastructure and plant or catering to new types of demand. It is in the planning of such new efforts that the biases toward optimism and strategic misrepresentation are likely to be largest. To be sure, choosing the right class of comparative past projects would become more difficult when planners are forecasting initiatives for which precedents are not easily found, for instance the introduction of new and unfamiliar technologies. However, most projects, and especially major ones, are both non-routine locally and use well-known technologies that have been tried out elsewhere. Such projects would, therefore, be particularly likely to benefit from the outside view.

Below, it is demonstrated how project managers and decision makers may use the outside view to perform due diligence of business cases, cost-benefit analyses, and social and environmental impact assessments for planned ventures.

## Potentials and Barriers for the Outside View

As argued in Flyvbjerg (2007a), two types of explanation best account for forecasting inaccuracy: optimism bias and strategic misrepresentation. The outside view was originally developed to mitigate optimism bias, but it may help mitigate any type of bias, including strategic bias, because the outside view bypasses such bias by cutting directly to empirical outcomes and building conclusions about future events on these. Even so, the potentials for and barriers to using the outside view will be different in situations where (1) optimism bias is the main cause of inaccuracy as compared to situations where (2) strategic misrepresentation is the chief reason for inaccuracy. We therefore need to distinguish between these two types of situation when endeavoring to apply the outside view in practice.

In the first type of situation--where optimism bias is the main cause of inaccuracy--we may assume that managers and forecasters are making genuine mistakes and have an interest in improving accuracy. Consider, for example, the students mentioned above, who were asked to estimate their future academic performance relative to their peers. We have no reason to believe that the students deliberately misrepresented their estimates, because they had no interest in doing so and were not exposed to pressures that would push them in that direction. The students made honest mistakes, which produced honest, if biased, numbers regarding performance. And, indeed, when students were asked to take into account outside-view information, we saw that the accuracy of their estimates improved significantly. In this type of situation--when forecasters are honestly trying to gauge the future--the potential for using the outside view to improve outcomes is good. Managers and forecasters will be welcoming the outside view, and barriers to using it will be low, because no one has reason to be against a methodology that will improve their forecasts.

In the second type of situation--where strategic misrepresentation is the main cause of inaccuracy--differences between estimated and actual costs and benefits are best explained by political and organizational pressures. Here managers and forecasters would still need the outside view if accuracy were to be improved, but managers and forecasters may not be interested in this because inaccuracy is deliberate. Biased forecasts serve strategic purposes that dominate the commitment to accuracy and truth. Consider, for example, the case of urban rail. Here, the assumption of innocence regarding estimates typically cannot be upheld. Cities compete fiercely for approval and for scarce national funds for such projects, and pressures are strong to present business cases as favorably as possible, that is, with low costs and high benefits, in order to beat the competition. There is no incentive for the individual city to debias its forecasts, but quite the opposite. Unless all other cities also debias, the individual city would lose out in the competition for funds. Managers are on record confirming that this is a common situation (Flyvbjerg and Cowi 2004:36-58). The result is the same as in the case of optimism: managers promote ventures that are unlikely to perform as promised. But the causes are different as are possible cures.

Under these circumstances the potential for using the outside view is low--the demand for accuracy is simply not present--and barriers are high. In order to lower barriers, and thus create room for the outside view, incentives must be aligned to reward accurate forecasts and punish inaccurate ones, for instance by allocating forecasting risk directly to investors so they will carefully scrutinize the accuracy of forecasts before deciding to invest in a project. Below, we will see an example of how this works, in the case study of the A-Train. Furthermore, to curb strategic forecasts governments and other funders may have to make the outside view mandatory and make funding contingent on the application of this particular methodology, as has been done in the UK and Denmark (UK Department for Transport 2006, Danish Ministry for Transport 2008). However, better methods alone will not solve the problem, because any method can be gamed (Winch and Maytorena 2011:354). Better methods need to go hand in hand with better incentives.[4] Finally, promoters' forecasts should be made subject to quality control and due diligence by other parties, including banks and independent bodies such as national auditors or independent analysts. Such bodies would need the outside view or similar methodology to do their work. Projects that were shown to have inflated benefit-cost ratios would not receive funding or would be placed on hold. The higher the stakes, and the higher the level of political and organizational pressures, the more pronounced will be the need for measures like these. In the next section we will see what quality control and due diligence based on the outside view might look like in practice.[5]

## What is Due Diligence?

Due diligence is a term used for the performance of an investigation of a business, an investment, or a person with a certain standard of care. Due diligence may be a legal obligation, but the term more commonly applies to voluntary investigations. Common examples of due diligence are the process through which a potential buyer evaluates a target company or its assets for acquisition or the process through which a potential investor evaluates a major investment for its costs and benefits. The theory behind due diligence holds that performing this type of investigation contributes significantly to informed decision making by enhancing the amount and quality of information available to decision makers and by ensuring that this information is systematically used to deliberate in a reflexive manner on the decision at hand and all its costs, benefits, and risks (Bing 2007, Chapman 2006).

Here, focus will be on the process through which a potential investor, public or private, may evaluate a major planned investment for its costs, benefits, and risks. To perform due diligence of the estimated costs and benefits for a planned project involves an evaluation of whether the cost and benefit forecasts in the project's business case are likely to be over- or underestimated and by how much. Like other forecasters, cost and benefit forecasters are faced with the following two basic questions:

1. What is the expected value of the variable being forecasted (here costs and benefits of the project in question)?

2. What is the variance of the expected value, i.e., the likelihood that the value will be, say, 5, 10 or 50 percent lower or higher than the expected value?

To perform due diligence of a given cost or benefit forecast – i.e., to evaluate whether the forecast is likely to be over- or underestimated – is equivalent to judging how well these two questions have been addressed by the forecasters. Below, it is demonstrated how this may be done in practice.

For a given cost or benefit forecast, the specific steps involved in performing due diligence based on the outside view are:

1. Identification and description of the business case or forecast to be evaluated.
2. Establishing a benchmark that represents the outside view, against which performance may be measured.
3. Using the benchmark to evaluate performance in the forecast in question.
4. Checking the forecaster's track record from other, similar, forecasts.
5. Identifying further cost and benefit risks.
6. Establishing the expected outcome.
7. Soliciting comments from the forecaster.
8. Concluding as to whether the forecast is over- or underestimated and by how much.

The outside view, as described above, is brought into due diligence by establishing the benchmark used in due diligence on the basis of performance in a group of projects similar to the one being investigated in the due diligence. Below it is shown how this is done in practice and we will see how each of the eight steps is carried out. Finally, we will discuss accuracy in cost and benefit forecasting and how it may be improved.

It should be emphasized that due diligence and quality control is not about eliminating uncertainty from business cases and forecasts. That is impossible. Uncertain events will always influence actual outcomes and thus the accuracy of forecasts. For instance, air fares change and influence rail patronage, or fuel prices change and affect demand for highways. This may be called *inherent uncertainty*. Nevertheless, much uncertainty – and especially bias – is not inherent but is caused by avoidable error and strategic behavior. Due diligence is aimed at reducing uncertainty by getting a clear picture of the size and types of uncertainty that apply to specific decisions and forecasts.

## Due Diligence in Practice

This section considers a case of typical state-of-the-art forecasting for a major capital project, namely a planned multibillion-dollar rail investment project. For reasons of data and presentation, the focus is on benefits and demand. However, the issues and methodology covered are general and apply to all types of costs and benefits and to social and environmental impacts. The project has been anonymized for reasons of confidentiality and is here called the "A-Train." The delivery model for the A-Train is a government-sponsored public-private partnership with a private concessionaire responsible for building and operating the service. This type of delivery model is today common for major infrastructure development in many parts of the world.

Demand forecasts are important to project managers and decision makers because they influence both cost and revenue risks of projects. Forecasts affect basic design decisions because they determine capacity and size of the facilities and systems being planned, for rail for instance number of tracks, number and length of trains, length of station platforms, and size of stations. An inflated passenger forecast for Bangkok's US$2 billion Skytrain resulted in a greatly oversized system, with less than half of the projected passengers, station platforms too long for the shortened trains, many trains and cars sitting in the garage, because there was no need for them, terminals too large, etc. (Flyvbjerg, Holm, and Buhl, 2005:132). Similarly, in areas of low population density, rail stations may require extensive car parking. If ridership forecasts are inflated, tens of thousands of parking spaces may be built that may not be needed, at a cost of tens of thousands of dollars per space. That would mean piling cost upon cost to meet fictional demand. The impact of the inflated forecast is that revenues will be lower than expected and costs higher than necessary, increasing viability risks. Given the high cost of major infrastructure projects, the irreversibility of decisions, and the limited availability of resources, it is crucial to focus on the quality of demand forecasts.

In public-private partnerships there is special focus on demand risk, because such risk typically needs to be allocated, first, between the public and private sectors, and second, between individual parties within those sectors, e.g., between local and national government and between several private investors. So, too, for the A-Train, where size and allocation of demand risk was a key issue. The question of whether the project's demand forecast may be over- or underestimated therefore became central to the project management process, and especially to parties who were asked to take on demand risk as part of their investment.

In what follows, due diligence is used to answer the question of whether the A-Train's demand forecast may be over- or underestimated. The due diligence was initiated by an investor who was invited to fund the A-Train and to take on a substantial part of the project's demand risk.

Again, the methodology used is general, although the data presented are particular to rail. For the method to work for other types of projects than rail, data for those types of projects would simply have to be substituted for the rail data in both the forecast to be evaluated and in the benchmark used to carry out the evaluation.

Each of the eight steps in due diligence listed above are covered in turn below.

## The Case of the A-Train

As part of step one – identification and description of the forecast to be evaluated – the sponsors' demand forecast for the A-Train is summarized in Table 1 and explained in the main text.

*Table 1: Sponsors' demand forecast for the A-Train, final business case. Predicted demand, downside case, and ramp up profile are shown as estimated by the sponsors' forecaster. See main text for discussion.*

| Item | Forecast |
|---|---|
| Predicted demand (expected outcome) | 14.1 million passengers, first year |
| | 17.7 million passengers, after 10 years |
| Downside case (variance, risk) | Standard deviation: 9.1% |
| | 95% confidence level: 15% passenger shortfall |
| | Likelihood of passenger shortfall of 15% or more: 5% |
| Ramp up profile (first five years after going live) | Year 1: 60% of forecast |
| | Year 2: 75% of forecast |
| | Year 3: 85% of forecast |
| | Year 4: 95% of forecast |
| | Year 5: 100% of forecast |

The sponsors' demand forecast, made for the A-Train final business case, predicted 14.1 million passengers for the opening year, growing to 17.7 million passengers per year ten years later. The sponsors and their forecaster considered this forecast to be the expected outcome taking into account the balance of probabilities of all the different risks and uncertainties in the modeling and forecasting process. Thus the sponsors considered the chance to be 50:50 for each of the events that the forecast would be exceeded and not be reached, respectively.

As an estimate of variance of the expected outcome, the sponsors presented what they called a "downside case" for demand and revenues, with 15 percent lower demand and revenues than the main forecast. The sponsors stated they had 95 percent confidence that these figures would be exceeded, which is equal to a 5 percent risk of a demand or revenue shortfall of 15 percent or more. This is equivalent to a standard deviation of forecasting outcomes of 9.1 percent. The downside case was produced by the sponsor's forecasters on the basis of their experience from Monte Carlo simulations of risk in other projects, according to the forecasters.

Finally, the sponsors of the A-Train estimated a five-year ramp up profile for demand, beginning with 60 percent of the forecast in Year 1 increasing to 100 percent in Year 5, as shown in

Table 1. Ramp up denotes a delay in demand take-up during the first months or years of a new service. The delay may arise due to the slow spread of knowledge about the service, inertia in changing travelers' behavior, initial operational hiccups, and the like. According to the forecasters, the ramp up profile was established on the basis of a limited amount of available data from other systems and a degree of professional judgment with the forecasters. The data claimed to support the ramp up profile was not provided with the forecast.

## Best or Average Practice as Benchmark?

Step two of due diligence is establishing a benchmark against which individual forecasts may be measured, including the A-Train forecast presented above. This is perhaps the most consequential step of the due diligence exercise. Benchmarks typically establish best, average, or median practice. In cost and benefit forecasting, *best practice* is equivalent to being on target, i.e., to produce a forecast with zero or little error, meaning a forecast that is neither too high nor too low. Although best practice is clearly an ideal to be strived for, in forecasting it is not a useful benchmark for due diligence. This is because benchmarking against best practice would give us only one piece of information, namely how far the forecast in question is from the ideal, i.e., from zero error. In this case we would obtain none of the distributional information that Kahneman and Tversky argue is crucial for understanding forecasting problems and getting forecasts right. For instance, using best practice we would not know how often best practice is actually achieved in forecasting and how far from best practice forecasts usually are. In what follows, the benchmark is therefore established on the basis of *average and median practice*, not best practice. As we will see, average and median practice in rail demand forecasting is far from best practice.

Measured by sample size, Flyvbjerg, Holm, and Buhl (2005) contains the most comprehensive study that has been published so far of accuracy in travel demand forecasting. The study is based on a sample of 210 transportation infrastructure projects of which 27 are rail projects. For 23 of the 27 rail projects demand forecasts were overestimated. Table 2 shows the main results from the study as regards rail, with and without 2 statistical outliers. Without statistical outliers actual demand is on average 51.4 percent lower than forecasted demand. This is equivalent to rail demand forecasts being overestimated by 106 percent on average. Just as important, standard deviations are high indicating a high risk of large forecasting errors.

*Table 2: Accuracy in 27 rail demand forecasts. Accuracy is measured as actual divided by forecast demand for the first year of operations. The standard deviation is measured for actual divided by forecast demand. (Source: Flyvbjerg et al. 2005).*

|  | **Average accuracy** | **Standard deviation** |
| --- | --- | --- |
| 27 rail projects | 0.61 | 0.52 |
| 25 rail projects, excluding statistical outliers | 0.49 | 0.28 |

For the present study, the 2005 sample was expanded to cover 475 transportation infrastructure projects of which 62 are rail projects that are comparable to the A-Train, in the sense they were built using technologies and under regulatory regimes that are comparable to those that apply to the A-Train. The dataset is the largest of its kind. All large transportation infrastructure projects, for which valid and reliable data on forecasting accuracy were available, have been included in the set. For a detailed description of the data, see Flyvbjerg (2005a) and Flyvbjerg, Holm, and Buhl (2005).

Table 3 shows results from the new, expanded study. For 53 of the 62 rail projects in the dataset, demand forecasts were overestimated. Provided that an estimate was high, the average accuracy of the demand estimate was 0.50 for these projects, that is, the overestimate was 100 percent on average and half the forecasted passengers never showed up on the trains. Accuracy in the new, larger sample of forecasts is approximately the same as in the earlier, smaller sample. Without statistical outliers, accuracy is 10 percentage points higher in the larger sample than in the smaller one, although accuracy is still low with actual demand being 41 percent lower than forecast demand on average. This is equivalent to rail demand forecasts being overestimated by 69 percent on average. Again standard deviations are high indicating high risks of large errors in rail demand forecasting.

*Table 3: Accuracy in 62 rail demand forecasts. Accuracy is measured as actual divided by forecast demand for the first year of operations. The standard deviation is measured for actual divided by forecast demand. (Source: Flyvbjerg database).*

|  | **Average accuracy** | **Standard deviation** |
| --- | --- | --- |
| 62 rail projects | 0.63 | 0.41 |
| 61 rail projects, excluding statistical outlier | 0.59 | 0.33 |

Table 4 shows the quartiles for the expanded study with and without the statistical outlier. The difference between including and excluding the outlier is small. In what follows the figures without the outlier are used, taking into account the possibility that the outlier may be caused by measurement error.

The table shows that for 25 percent of rail projects actual demand is 65 percent lower or more than forecast demand, equal to an overestimate in forecasts of 186 percent or more in a quarter of cases. For half the projects in the sample actual demand is 49 percent lower or more than that forecast, equal to a overestimate of 96 percent or more. Finally, the table shows that for 75 percent of rail projects actual demand is 22 percent lower or more than forecast, equal to an overestimate in demand forecasts of 28 percent or more for these projects.

The numbers show that the risk of overestimates in rail demand forecasts is high. For instance the risk is 50:50 of an overestimate of 96 percent or more. It is the data for these 62 rail demand forecasts that are used as a benchmark against which other rail demand forecasts may be measured, including the A-Train forecast.

*Table 4: Quartiles for accuracy in 62 rail demand forecasts. Accuracy is measured as actual divided by forecast demand for the first year of operations (source: Flyvbjerg database).*

|  | **25%** | **50%** | **75%** |
|---|---|---|---|
| 62 rail projects | 0.36 | 0.54 | 0.79 |
| 61 rail projects, excluding statistical outlier | 0.35 | 0.51 | 0.78 |

## Applying the Outside View

Step three in due diligence based on the outside view is to employ the benchmark to evaluate performance for the forecast in question. As mentioned above, research by Daniel Kahneman, Amos Tversky, and others has established that a major source of error in forecasts is a prevalent tendency with forecasters to underweigh or ignore variance of outcomes and thus risk. This is to say that not only do forecasters err in estimating the expected outcome of the variable they are forecasting. More often than not they make even larger errors in estimating the variance (typically measured in standard deviations) of the expected outcome. Here we therefore emphasize how to benchmark variance in a business case forecast, again with rail and the A-Train demand forecast as examples of the general problematic.

Table 5 compares the A-Train downside case with the benchmark established above. The table shows that:

1. The standard deviation of accuracy in the benchmark is 33 percent, compared with 9.1 percent in the A-Train downside case.

2. The 95 percent confidence level is equivalent to a demand shortfall of 85 percent in the benchmark, compared with 15 percent in the downside case.

3. The likelihood of a demand shortfall of 15 percent or more is 80 percent in the benchmark, compared with 5 percent in the downside case.

*Table 5: Main results from benchmarking of A-Train downside case against outcomes of 61 actual rail demand forecasts.*

|  | **A-Train downside case** | **Benchmark (61 forecasts)** |
|---|---|---|
| Standard deviation | 9.1% | 33% |
| 95% confidence level | 15% shortfall | 85% shortfall |
| Likelihood of demand shortfall of 15% or more | 5% | 80% |

Compared with the benchmark, the standard deviation of the A-Train demand forecast appears to be highly underestimated. As mentioned, this is a common error in forecasting and it results in forecasts that are optimistic in the sense that the actual range of possible outcomes, i.e., actual uncertainty, is larger than the estimated range of possible outcomes, i.e., estimated uncertainty. The A-Train demand forecast is highly optimistic in this sense. To illustrate, the risk is a full 16 times higher in the benchmark than in the A-Train demand forecast of a demand shortfall of 15 percent or more.

## Dealing With Demand Ramp Up

For a demand forecast, step 3 also includes evaluating the ramp up of patronage. Ramp up denotes the delay in demand take-up during the first months or years of a new service, as mentioned above. The accuracy of ramp up is crucial to realizing forecasted revenues. Getting ramp up wrong has been known to seriously jeopardize the viability of projects. The practice in business cases and cost-benefit analyses of discounting future benefits and costs to net present values makes the accuracy of ramp up even more important, because first-year revenues and benefits are attached the greatest weight in such discounting. Furthermore, public and media scrutiny will also be at its highest during the first period of operation, i.e., the ramp up period, making accurate estimates of ramp up important to good public relations and corporate reputation.

Table 6 compares estimated ramp up for the A-Train with actual ramp up in the 11 rail projects for which comparable ramp up data was available from the dataset. The 60 percent starting point assumed for the A-Train ramp up profile in year one compares to 41 percent actually realized for the 11 projects for which data are available. This is an overestimate for the A-Train of 46 percent for year one, compared with the benchmark. For the ramp up profile itself, the A-Train forecast assumes an increase of 40 percentage points over five years, from 60 to 100 percent, compared to 14 percentage points actually realized in the benchmark, from 41 to 55, which amounts to an overestimate by a factor 2.86, compared with the benchmark.

Table 6 shows that five years after going live, actual demand in the 11 projects was approximately half of the forecasts. The data thus does not support the estimated ramp up for the A-Train. The table shows that with 45 percent of demand having not materialized after five years of operations it hardly makes sense to even speak of ramp up for the 11 projects. Rather, we have a case of consistent underperformance as regards later-year demand, if first-year demand turned out substantially lower than forecasted. This conclusion holds both for ramp up for each of the 11 projects and for average ramp up for all 11 projects.

Four published studies on ramp up in rail and road schemes support this conclusion and thus do not support the A-Train ramp up profile (Flyvbjerg 2005a, Pickrell 1990, Standard and Poor's 2005, UK National Audit Office 2004). All four studies found that error and bias in demand forecasting is not caused by forecasters not considering ramp up but by actual ramping up – if it exists at all – being far less pronounced and taking many more years than assumed by forecasters. Beyond ramp up, many projects still fail to meet demand forecasts. The studies even documented negative ramp up for a number of projects, i.e., actual demand dropped over time.

Based on this evidence, most likely the A-Train demand forecast is overestimated as regards ramp up. The available evidence indicate that this is common for rail demand forecasts and is a major source of bias in such forecasts.

As mentioned above, the data claimed to support the A-Train ramp up profile was not provided with the forecast. Nor were the data available upon request, even to a potential investor, who was invited to invest substantially in the project and to take on a significant part of the demand and ramp-up risks. This was seen as indication that the data did not exist but were simply postulated by the forecaster, which undermined trust in the forecast and the forecaster.

*Table 6: Estimated ramp up profile for the A-Train and actual average ramp up for 11 other rail projects. Actual demand in percent of forecast demand. See main text for detailed explanation.*

| Year | A-Train (planned) | 11 rail projects (actual) |
|---|---|---|
| 1 | 60 | 41 |
| 2 | 75 | 49 |
| 3 | 85 | 68 |
| 4 | 95 | 51 |
| 5 | 100 | 55 |

# Checking the Forecaster's Track Record

Step four in the method for due diligence followed here consists in checking the forecaster's track record from other, similar, forecasts. Forecasters typically attempt to lend credibility to their bid to do a forecast for a specific project by listing forecasting experience from other similar projects. But rarely do forecasters provide ex post data on actual accuracy of forecasts made in the past. If asked to provide such evidence of actual track record, forecasters typically reply that the evidence is confidential and cannot be released (a common explanation for private projects) or that the evidence is not available, because they have done no systematic follow-up studies of accuracy, and neither have others (a common explanation for public projects). Either way, evidence of documented track record of accuracy of past predictions is typically not available from forecasters.

So, too, for the A-Train forecaster, which is a large consultancy with global presence and extensive experience in demand forecasting. For the A-Train forecast, they listed credentials from more than 20 previous, similar forecasts they had done. But when asked for evidence of track record for the accuracy of these forecasts, no evidence was provided. The reasons given were the same as above, i.e., no ex post data had been collected or such data were confidential. The party asking for evidence was a potential major investor, who would take on substantial demand risk, should they decide to invest in the A-Train. The forecaster's inability or unwillingness to provide evidence of the actual accuracy of their forecasts therefore meant, in effect, that the investor was asked to commit hundreds of millions of dollars to the A-Train venture on sheer belief and trust that the demand forecast would be accurate. But "Trust me!" has historically not been the best basis for handing over large sums of money to strangers to invest, which is why due diligence is important, of course.

For the A-Train, the evaluation above shows that such trust would clearly be unwarranted. Moreover – and unfortunately for the forecaster – the dataset used here to establish the benchmark for the due diligence contains information about track record for two of the rail projects mentioned by the A-Train forecaster on their credentials list. These two projects, each a multibillion-dollar venture, performed as shown in Table 7.

*Table 7: Demand overestimate in percent for two rail forecasts done by the A-Train forecaster. Each figure shows by how many percent that variable was overestimated. Mode share is the percentage of total travellers expected to use rail.*

| Project | First-year demand overestimate, % | Later-year demand overestimate, % | Mode share for rail overestimate, % | Comment by National Audit Office |
|---|---|---|---|---|
| **Project 1** | 500 | 250 (4th year) | 350 | n.a. |
| **Project 2** | 175 | 150 (5th year) | n.a. | "completely failed ... grossly overestimated" |

Table 7 shows that the forecaster's track record for the two projects is significantly poorer than the benchmark, which is already quite poor with a median overestimate of 96 percent (see above).

The national audit office in the country where Project 2 is located decided to commission an in-depth ex-post study of benefits and costs of the project, when demand shortfalls proved the project financially nonviable and a national scandal ensued leaving high-ranking politicians and officials with egg on their faces. The study found that the demand forecasts "completely failed to capture development in the [rail passenger] market at [the location of the rail project]". The study also found that, "the private consortium [bidding for the link] was presented with grossly overestimated [demand] forecasts, which was [later] criticized [by the National Audit Office] but without response [from the forecaster]." Thus the private consortium that funded Project 2 was misled by the demand forecast, and the consortium – to their later grief – did not carry out due diligence for the project, like the one described here for the A-Train. As a result the project went bankrupt and the consortium lost a fortune. When the National Audit Office criticized the forecaster for this, the forecaster simply ignored the critique and did not bother to answer the auditor, according to the audit.

In the due diligence of the A-Train demand forecast, trust in the A-Train forecaster and forecast suffered from this combination of:

1. The forecaster's unwillingness or inability to document accuracy for their previous forecasts.

2. The forecaster's extremely poor track record for the few projects for which forecasting accuracy could be established.

3. The forecaster being discredited in public by a national audit office for having "completely failed" and not even responding to the critique when invited.

In sum, for the A-Train, the available information regarding the forecaster's track record does not support the forecaster's claim that the A-Train demand forecast would be accurate. The benchmark documents an expected demand overestimate of 69 percent (median 96 percent) and there are no a priori reasons to assume that the expected overestimate in the A-Train forecast would be lower than in the benchmark. Quite the opposite, given the fact that for the two projects on the forecaster's credentials list for which we have data, the average overestimate was 338 percent for the first-year estimate – i.e., 4.9 times the already dismal benchmark – and 200 percent for the four-to-five-year estimate. Given the available information it is prudent to assume a very high risk of overestimate in the A-Train demand forecast and thus a very high risk of revenue shortfalls. Prospective investors would be well advised to require more information about the track record of the A-Train forecaster before committing funds to this project, especially if investors were asked to carry all or a substantial part of the demand risk.

In a wider perspective, forecasters' unwillingness or inability to document the accuracy of their forecasts runs directly against much-touted policies for evidence-based decision making and management and it amounts to forecasters effectively asking their clients to buy – both in the sense of believing in and paying for – their product without evidence to substantiate that the product actually works, or has worked in the past. The type of forecasting we are studying here is globally a billion-dollar business. Very few other businesses of this size would get away with not demonstrating the track record and quality of their product. The point is that the forecasting business should also not be allowed to do this, and there are no defensible reasons they are. It is the task of quality control and due diligence to point to this problem and to make sure the track record is documented to the extent possible, as done above.

## Further Demand Risks

Step five investigates whether further demand risks exist, on top of those already described and analyzed above. A number of circumstances specific to the A-Train pose such further risks. Other things being equal, this indicates that demand shortfalls for the A-Train are likely to be higher than average shortfalls in the benchmark. The additional demand risks for the A-Train are:

1. Urban development in the catchment area for the A-Train is dispersed and low-density. Predicted demand for the rail line is partly dependent on increased density around stations. Changes in density and land use that were expected but did not materialize are a common cause of failed demand forecasts (Transportation Research Board 2007).

2. Part of the A-Train system runs a service to an airport located at the edge of a metropolitan area not far from the city's business center. Research shows that this type of location indicates a higher-than-average risk for lower-than-forecasted rail use, whereas higher distances from airports to the city edge and city center indicate higher rail use (Bradley 2005).

3. The planned fare level of the A-Train airport link seems high when compared to fare levels at other airport rail links. Thus a risk may exist that high fares could result in lower-than-forecasted demand. This risk is mitigated, however, by the fact that demand on airport rail links tend to be relatively inelastic to changes in fares, probably because air passengers generally have a high value of time (Bradley 2005, 9). To assess risk on this point would be to assess whether or not the high fares for the A-Train airport link are likely to fall within the area of inelasticity. Such assessment is pertinent but was not carried out by the forecaster.

4. Total A-Train revenues are particularly sensitive to demand risk on the airport link, because although the airport link is forecasted to carry only 10-14 percent of a total A-Train passengers, it is forecasted to generate 35-42 percent of total revenues. Thus if the airport link demand forecast is overestimated then this might significantly and negatively impact revenue for the total A-Train.

5. The database used for the present study includes data on forecasting accuracy for four airport rail links, two of which were forecast by the A-Train forecaster. The average overestimate for the four forecasts were 147 percent, which is more than double the benchmark. Thus the risk of demand overestimate and shortfall seems higher for airport links than for other rail links, given the available evidence.

6. Further demand risks exist that cannot be revealed, because this might disclose the identity of the A-Train. These risks are high and have been proved significant through statistical testing.

In sum, further demand risks for the A-Train indicate that demand shortfalls for the A-Train are likely to be higher than average shortfalls in the benchmark, which are already high.

## Establishing Expected Outcome

Step six establishes the expected outcome of the forecast. Given the optimism documented above in the A-Train demand forecast regarding variance, ramp up, etc., it appears prudent to point out that if the estimate of expected demand is as optimistic as the estimates of variance and ramp up then expected demand is overestimated. The large overestimates in the track record of the forecaster and the specific demand risks for the A-Train mentioned above also point in the direction of substantial risk of overestimation.

If the A-Train demand forecast is argued to be neither more or less accurate than the average rail demand forecast, then one would expect an overestimate of 69 percent as shown above for the benchmark. Expected actual demand would then be 59 percent of forecast demand, or *8.3 million first-year passengers instead of the 14.1 million predicted*. Thus 5.8 million forecasted passengers would not show up on the trains, resulting in large overcapacity and revenue shortfalls. With 90 percent confidence, actual demand would lie in the interval from 15 percent to 110 percent of forecast; with 80 percent confidence, in the interval from 23 percent to 101 percent of forecast; and with 50 percent confidence in the interval from 35 percent to 78 percent of forecast (see Table 8).

Given the evidence presented above, most likely the A-Train demand forecast is less accurate than the average rail demand forecast. Demand shortfalls are thus likely to be larger than those shown in Table 8.

*Table 8: Expected outcome and confidence intervals for A-Train demand, provided the A-Train forecast is neither more nor less accurate than demand forecasts in the benchmark.*

|  | **Actual/forecast demand** | **First-year passengers, mio** |
|---|---|---|
| Expected outcome | 0.59 | 8.3 |
| With 90% confidence | 0.15-1.10 | 2.1-15.5 |
| With 80% confidence | 0.23-1.01 | 3.2-14.2 |
| With 50% confidence | 0.35-0.78 | 4.9-11.0 |

## Do Private Projects Perform Better than Public Ones?

Step seven solicits comments from the forecaster regarding the previous six steps of the due diligence. When the A-Train forecasters were given the opportunity to comment on the methodology and results presented above they made the following two claims:

1. The accuracy of demand forecasting for bidding consortia and financial institutions has improved since the mid 1990's as more privately funded schemes, such as the A-Train, have come to the market. Private funding is likely to curb bias in forecasting and result in more accurate forecasts. The A-Train demand forecast is therefore believed to be more accurate than the forecasts in the benchmark, according to the A-Train forecaster.

2. The majority of transportation infrastructure schemes – including the A-Train – has unique characteristics. Therefore it is wrong to compare them to other projects, for instance comparing the A-Train project to the rail projects in the benchmark, as done above.

The first claim is covered in this section, the second in the next. However, for both claims it may be observed that the A-Train forecaster presented no evidence to support the claims, just as the forecaster presented no evidence to support the validity of the A-Train forecast itself. Again we therefore have to look elsewhere for evidence.

Table 9 shows the ratio of actual to forecast demand for privately funded rail projects in the database, built from the mid 1990's. The ratio of actual to forecast demand is 0.30 for privately funded schemes and 0.62 for public schemes, i.e., on average 30 percent of demand materialized for private projects versus 62 percent for public ones. Thus inaccuracy (bias) is more than twice as high for private schemes than for public ones. Although the number of private projects for which data are available is small, clearly the data in the table do not support the A-Train forecaster's claims regarding higher forecasting accuracy for privately funded schemes. In fact, the data supports the opposite conclusion, namely that demand forecasts for privately funded projects are more overestimated than for publicly funded ones.

This conclusion is supported by a large-sample study of toll roads carried out by Standard and Poor's (2004), which found that demand forecasts for privately funded roads were significantly more inaccurate (overestimated) than demand forecasts for public roads. Private ventures were found to be incentivized to inflate demand forecasts to make projects look good on paper in order to secure investor interest.

Two of the five privately funded projects in Table 9 are the two projects mentioned above under track record, for which the A-Train forecaster had themselves done the forecast, and for which first-year demand was overestimated by 175 and 500 percent, respectively. Not only were these forecasts more inaccurate than public-sector forecasts, they were also more inaccurate on average than the other private-sector forecasts presented in Table 9. Thus the A-Train forecaster's forecasts were the most inaccurate of all in the sample, and especially for private-sector projects. On this background, the forecaster's claim is undermined that the A-Train demand forecast is likely to be more accurate than the forecasts in the benchmark, because it is a forecast for a privately funded project, and private funding is likely to result in more accurate forecasts. It was concluded that either (a) the A-Train forecaster did not know the accuracy of their own forecasts, or (b) they knew and misinformed about it. Either way, these unfounded claims, contradicted by the data, eroded trust in the forecaster and their forecast.

For the above reasons, the forecaster's claim was rejected that because the A-Train involves private funding the A-Train demand forecast is likely to be more accurate than the forecasts in the benchmark.

*Table 9: Accuracy of demand forecasts for privately funded rail projects, first-year demand.*

| Project Type | Actual/forecast demand |
|---|---|
| 5 private projects | 0.30 |
| 56 public projects | 0.62 |
| All projects (61) | 0.59 |

# Are Projects Unique?

The second claim made by the A-Train forecasters under step seven of the due diligence was that projects are unique and therefore cannot be compared, including to benchmarks. This is a claim often made by forecasters and project managers. Nevertheless, the claim is wrong.[6]

It is correct that infrastructure projects typically have physical, locational, or other characteristics that are unique. The claim is incorrect, however, in the sense that research shows, with a very high level of statistical significance, that despite physical or other uniqueness rail and other infrastructure schemes are statistically similar in the sense that they are highly likely to have large demand shortfalls and large cost overruns (Flyvbjerg 2007). Statistical similarity is all that is needed for statistical prediction, and statistical prediction based on the outside view is an effective means to improve accuracy in forecasts as demonstrated by Flyvbjerg (2006).

This is the approach taken by the UK Treasury (2003, 1), when they observe about project appraisal:

> "There is a demonstrated, systematic, tendency for project appraisers to be overly optimistic. To redress this tendency appraisers should make explicit, empirically based adjustments to the estimates of a project's costs, benefits, and duration ... it is recommended that these adjustments be based on data from past projects or similar projects elsewhere."

"Similar" here means statistically similar and although the type of adjustments recommended by the UK Treasury were neglected by the A-Train forecaster such adjustments appear to be exactly what would be needed to produce a more reliable demand forecast for the A-Train. The "data from past projects and similar projects elsewhere," mentioned by the UK Treasury, are data like those presented above in the benchmark. The use of such data has been made mandatory in the UK by the UK Treasury and the UK Department for Transport (2006) to adjust for optimism bias in UK transportation cost forecasts following the methodology described in Flyvbjerg (2006). The Danish and Swiss governments have implemented similar measures to curb unrealistic forecasts (Danish Ministry for Transport 2008, Swiss Association of Road and Transportation Experts 2006).

# Final Step of Due Diligence: Are Forecasted Benefits Likely to Happen?

The eighth, and final, step in a due diligence of a cost or benefit forecast is to answer the question of whether the specific forecast is likely to be over- or underestimated, and by how much. In demand forecasting, such assessment is crucial, because demand impacts revenues, which directly impact viability for a given project. Demand also impacts costs because demand determines the size of facilities and systems, as explained above. Finally, demand impacts the environment and thus the results of environmental impact assessments.

For the A-Train demand forecast, the answer to this question is – based on the evidence presented above – that the forecast is likely to be overestimated. The central arguments for a likely overestimate are:

1. Average practice in a benchmark of 61 comparable passenger demand forecasts, taken from the world's largest database of such forecasts, is an overestimate of demand of 69 percent (median 96 percent). This is the expected overestimate of a demand forecast of this kind, unless there is evidence that the forecaster is likely to have been more accurate than the forecasters who did the forecasts in the benchmark. For the A-Train forecast such evidence does not exist, but quite the opposite, as shown above.

2. Variance of the expected outcome is significantly underestimated in the A-Train demand forecast. For instance, the risk of a demand shortfall of 15 percent or more is 16 times higher in the empirical benchmark than in the A-Train forecast.

3. Assumptions about ramp up in the A-Train forecast is unsupported by evidence and appears to be overestimated by approximately a factor 3.

4. The forecaster provides no track record to document accuracy in their previous forecasts. Evidence from previous forecasts done by the forecaster, for which data on accuracy could be found, show a high degree of overestimate, significantly higher than in the benchmark, which is already high.

5. The mode share for the A-Train airport link appears to be overestimated, that is, the percentage of total travellers expected to use rail, against the percentage expected to use car, bus, etc.

6. Issues of density, urban form, location, fare levels, and policy pose risks to demand that are larger for the A-Train than for the average project in the benchmark.

7. The forecaster's argument that demand forecasts in privately funded schemes, such as the A-Train, will be more accurate than in publicly funded ones, is not supported by evidence. In fact, for privately funded schemes for which data are available, demand overestimates are more than twice as high as in other schemes.

On balance, one cannot argue that the A-Train demand forecast is likely to be less overestimated than the average rail demand forecast in the benchmark. The opposite is more likely, given the available evidence, i.e., that the A-Train demand forecast is more overestimated than the benchmark, which has an average overestimate of 69 percent (median 96 percent).

However, if one gives the benefit of doubt to the A-Train demand forecast and assumes that it is neither more nor less accurate than the average rail demand forecast, then one would predict the following:

1. An overestimate of 69 percent, equivalent to actual demand being 59 percent of that forecasted, that is, 8.3 million passengers during the first year of services, instead of the forecaster's estimate of 14.1 million passengers. Thus 5.8 million of the forecasted passengers would most likely never show up on the train service
2. A 50 percent chance that actual demand will fall in the interval of 35 percent to 78 percent of the forecast, i.e., 4.9-11.0 million passengers

For the A-Train, most likely both the overestimate and the shortfall will be larger than these figures, given the evidence presented above.

In addition to the forecaster's numbers being a cause of concern, the forecaster's behavior also undermined trust when the forecaster:

1. Was unwilling or unable, upon request, to document accuracy of their previous forecasts.
2. Was unwilling or unable, upon request, to provide documentation for their assumptions regarding ramp up and instead justified these by "professional judgment," which was later proven optimistic by actual data.
3. Argued without evidence that because the A-Train is privately funded its demand forecast is more likely to be accurate, when evidence indicates that the exact opposite is the case, including for the forecasters own previous forecasts.
4. Argued without evidence that the A-Train is unique and may therefore not be compared to other projects, when evidence shows that such comparisons are justified and are mandatory elsewhere, including for UK government-funded projects.
5. Had their forecasts for a multibillion-dollar high-speed rail project, which was similar to the A-Train, and which went bankrupt, thoroughly debunked by a national audit office and did not respond to the critique when asked to do so.

On this basis it was concluded that potential investors in the A-Train – public or private – would be well advised to trust neither the forecaster nor the forecast and to not take on demand risks until demand and revenues had been reassessed in the manner described above. This advice was followed by the investor who commissioned the due diligence presented here. *The investor decided not to invest in the A-Train*.

## Fire the Forecaster?

Two observations may be made about the A-Train demand forecast, as described and assessed above:

1. This is a normal, state-of-the-art demand forecast, like thousands of such forecasts made every year at substantial cost to clients.[7]

2. The forecast is garbage, of the type garbage-in-garbage-out. In fact, the forecast can be argued to be worse than garbage, because it gives the client, investors, and others who use the forecast, the impression that they are being informed about future demand, when, in fact, they are being misinformed. Instead of reducing risk in decision making, forecasts like this increase risk by systematically misleading decision makers and investors about the real risks involved.

The latter was the point made by the national audit study mentioned above about another forecast done by the A-Train forecaster for another multibillion-dollar high-speed rail project. The national auditor called the forecast "completely failed" and "grossly overestimated" and documented how the forecast had misled investors to commit billions of dollars to a project that proved non-viable, became a national scandal, and had to go through painful financial restructuring. The risk is very real that the A-Train faces a similar fate, but the A-Train demand forecast hides this for the time being by its optimism, documented above.

First, if the promoter who commissioned and paid for the original A-Train demand forecast requested a valid and reliable forecast, then the promoter should ask to have their money back on the grounds that the forecaster has delivered a faulty product. In this case, the promoter may also consider to later seek compensation from the forecaster for possible damages caused by the faulty forecast. Second, if the promoter and forecaster conspired to knowingly produce an inflated forecast in order to lure investors to fund the A-Train project by making it look good on paper, then investors may consider seeking compensation for damages caused by the conspiracy.

In general, risks for major project investments like the A-Train are at least as large, as misunderstood, and as mismanaged as risks in the financial markets, and probably more so because data about project risks are less available than are data about market risks. In finance, biased forecasts and bad risk assessments led to the 2007-2009 market collapse, which exposed financial forecasters' dubious practices. As a consequence, economists began to discuss the necessity of "firing the forecaster" (Akerlof and Shiller 2009, 146). This is prudent advice, also for major project investment decisions. But here, firing the forecaster may be letting off forecasters too easily. Some forecasts are so grossly misrepresented and have such dire consequences that we need to consider not only firing the forecasters but suing them, too, for the losses they incur. In a few cases, where forecasters foreseeably produce deceptive forecasts, criminal penalties may be in place, if the damage caused by forecasters is severe enough. So far, no forecaster has been sued for an inaccurate forecast, to the best knowledge of the present author. This may well be changing with an AUD$150 million class action suit under way in Australia, against the traffic and revenue forecaster for the Clem Jones toll road in Brisbane for producing "woefully inaccurate" forecasts. The suit is organized on behalf of 700 investors who lost their money when the toll road went bankrupt (Samuel 2011, Wilson 2012). The suit is historic as the first of its kind and should therefore be studied carefully by forecasters and forecasting scholars alike.

Taleb (2010:163) says about deceptive forecasters:

> "People ... who forecast simply because 'that's my job,' knowing pretty well that their forecast is ineffectual, are not what I would call ethical. What they do is no different from repeating lies simply because 'it's my job.' Anyone who causes harm by forecasting should be treated as either a fool or a liar. Some forecasters cause more damage to society than criminals." (Taleb 2010, 163).

Management research has recently developed strong theory to help identify the "fools" and "liars" Taleb talks about here. Being academic, the theories use more polite and opaque language, needless to say, describing Taleb's fools as people who are subject to "optimism bias" and the "planning fallacy" and liars as those practicing "strategic misrepresentation," "agency," and the "conspiracy of optimism" (Flyvbjerg, Garbuio, and Lovallo 2009; Gardener and Moffat 2008; Hirt 1994; Weston et al. 2007).[8] This research has demonstrated that fools and liars constitute the majority of forecasters and that forecasts are used as sinister tools of fraud more often than we would like to think. Wachs (1990:146, 1986:28) found "nearly universal abuse" of forecasting in this manner, and it is common in all sectors of the economy where forecasting routinely plays an important role in policy debates, according to Wachs. Forecasts are used to "cloak advocacy in the guise of scientific or technical rationality," in Wachs' (1989, 477) classic formulation, and there is no evidence that things have improved since Wachs did his pioneering studies (Flyvbjerg 2007a).

However, recent research has also developed the concepts, tools, and incentives that may help curb delusional and deceptive forecasts. The outside view and due diligence, presented above, was developed with this purpose in mind. So was reference class forecasting (Flyvbjerg, Holm, and Buhl 2005; Flyvbjerg 2006), which has been officially endorsed by the American Planning Association (2005) and made mandatory by government in the UK, Denmark, and Switzerland (UK Department for Transport 2006, Danish Ministry for Transport 2008, Swiss Association of Road and Transportation Experts 2006). Finally, the incentive structures designed by Flyvbjerg, Bruzelius, and Rothengatter (2003) and Flyvbjerg (2007a) are means to curb both optimism bias and strategic misrepresentation in forecasting. Given the knowledge and tools we have today for making better forecasts, there is little excuse for accepting forecasts like that for the A-Train, which is incompetent at best (made by fools) and fraudulent at worst (made by liars). We have the knowledge and tools to disclose the incompetence and fraud and should clearly do so, given the large financial, economic, social, and environmental risks that are at stake in this type of multibillion-dollar investments. Luckily, government and business are beginning to see this.

Forecasting fools should be sent back to school. They make errors unknowingly and are therefore likely to be motivated to do things differently, once the errors are pointed out to them and better ways of forecasting are presented. When in school, they would need to unlearn conventional forecasting methods and to instead learn about optimism bias, the planning fallacy, the inside view, black swans, strategic misrepresentation, political power, principal-agent problems, the outside view, quality control, due diligence, reference class forecasting, incentives, and accountability. Their key teachers should be people like Daniel Kahneman, Nassim Taleb, and Martin Wachs. The main take-away for students would be a realization that a fool with a tool is still a fool, and that, therefore, they must stop being fools.

Forecasting frauds should be fired and possibly prosecuted, depending on how serious their fraud is. Forecasting frauds knowingly bias their forecasts for personal or organizational gain, for instance to obtain approval and funding for their projects. Forecasting frauds are therefore not immediately motivated to change their ways. Fraud is their business model. This is why carrots will be ineffective with frauds and sticks must unfortunately be brought out. A light stick may be found in professional ethics. Professional organizations like PMI, APM, APA, RTPI, and other professional societies for managers, planners, engineers, and economists could use their codes of ethics to penalize, and possibly exclude, members who do unethical forecasts. Given how widespread unethical forecasts are, it is interesting to note that such penalties seem to be non-existent or very rare, despite codes of conduct that explicitly state that misinforming clients, government, citizens, and others is unacceptable behavior by members of these organizations. By not sanctioning fraudulent forecasts, and thus quietly accepting them, professional organizations become co-responsible for their existence. This needs to be debated openly within the relevant professional organizations. Malpractice in project management should be taken as seriously as malpractice in other professions, like medicine and law. To not take malpractice seriously amounts to not taking the profession of project management seriously.

A heavier stick would be to make forecasters financially liable for inaccurate forecasts and to reward them for accurate ones, something that is rarely done today, but which would undoubtedly improve accuracy. If there are no incentives to get forecasts right, perhaps we should not be surprised when they are wrong. Finally, an even heavier stick would be to prosecute forecasters who, through blatant negligence or deliberate deception, cause serious damage to others with their forecasts, be they government, corporations, or citizens. Here the Sarbanes-Oxley Act-like legislation, which has recently been implemented in many countries around the world, signifies progress (United States Congress 2002). Such legislation is aimed at corporate fraud and in many contexts fraudulent forecasts may be considered part of corporate fraud. Forecasters are therefore no longer as safe from prosecution as they were earlier (Bainbridge 2007). This in itself should have a sobering effect on forecasters, if not before then when someone has the good sense to sue a fraudulent forecaster.

## Areas for Further Research

The present paper has taken a first step in researching a much-neglected problem in scholarly work on project management, namely what to do about the fact that front-end estimates of costs and benefits – used in the business cases, cost-benefit analyses, and social and environmental impact assessments that typically support decisions on projects – are often significantly different from actual ex post costs and benefits. Such estimates are therefore commonly poor predictors of the actual value and viability of projects and estimates frequently misinform decision makers on projects instead of informing them. The paper offers (1) a new methodology for quality control of front-end estimates, based on theories of the planning fallacy and the outside view, (2) a case study with empirical test and demonstration of the proposed methodology, and (3) incentives for getting front-end estimates right, including sanctioning forecasters who cause damage to others with their forecasts. Further research is needed in the following areas:

1. Case studies with additional examples of how the proposed (or similar) methodology for quality control works in practice, akin to the case study presented above, with a view to obtaining more in-depth knowledge and a wider variety of experience regarding the strengths and weaknesses of the methodology and the theories behind it. Studies should cover the full range of outcome variables and not just demand as in the case study above.
2. Statistical analyses of a sample of projects that have been subjected to the methodology for quality control (or alike methodology), with the specific purpose of measuring whether estimates become more accurate and actual outcomes improve with the application of the methodology. The sample should be large enough to allow tests for statistical significance. For valid results, estimates and actual performance should be measured consistently in the same manner across projects using standard methodology, something that is rarely done today in research on project performance. Such statistical analyses would inform further development of both the theory and methodology of quality control and due diligence in project management.
3. Studies of how different types of incentives and disincentives impact accuracy of front-end estimates, including the threat of legal action against forecasters who cause serious damage with their forecasts. Australia's Clem Jones class action suit, mentioned above, is likely to be an interesting case study in this context. Impacts of changed or new incentives should be studied not just for individual forecasters but across the wider forecasting community.

# Notes

[1] The author wishes to thank Daniel Kahneman for generously taking the time to discuss in several meetings and correspondence the idea of quality control and due diligence of decision making, based on theories of the planning fallacy and the outside view. The author also wishes to thank Alexander Budzier, Ole Jonny Klakegg, Jens Krause, Jürgen Laartz, Petter Næss, and Martin Wachs for valuable comments on earlier versions of this paper. Research for the paper was supported in part by The Danish Council for Strategic Research.

[2] This is assuming that the project has not fallen victim to lock-in at an earlier stage and that promoters are genuinely interested in understanding outcomes, which is not always the case, for instance where promoters have locked in on a certain design or construction interests dominate operations interests (Cantarelli et al. 2010).

[3] Buehler, Griffin, and Ross (1994) originally used the term "planning fallacy" to explain error and bias in schedule estimates. Later, the term has been used to also explain error and bias in cost and benefit estimates, in addition to in schedule estimates.

[4] The wider issue of the politics of business case development are detailed in Flyvbjerg, Bruzelius, and Rothengatter (2003) and Flyvbjerg, Holm, and Buhl (2005), including the design and implementation of incentive structures that would make decision making well-informed and accountable.

[5] The existence of strategic misrepresentation does not exclude the simultaneous existence of optimism bias, and vice versa. In fact, it is realistic to expect such co-existence in forecasting in large and complex ventures and organizations. This again underscores the point that improved methods and measures of accountability must go hand in hand if the attempt to arrive at more accurate forecasts is to be effective.

[6] Recent research on major ICT projects shows that one of the most reliable predictors of larger-than-normal cost overrun is project management seeing their project as unique (Budzier and Flyvbjerg 2011, Flyvbjerg and Budzier 2011). This is easy to understand. Managers who see their projects as unique believe they have little to learn from other projects. They therefore accumulate less knowledge and experience to inform their decisions. This, in turn, may prove ruinous, especially in analyzing and managing risk, which, per definition, requires knowledge of variation across projects and thus a view of projects as non-unique, at least in a statistical sense.

[7] Most likely, the forecast will prove to be less accurate than the benchmark, as explained above, but the forecast is unlikely to be a statistical outlier.

[8] A grey area exists between deliberate misrepresentation and innocent optimism. Forecasters may know, or have a hunch, that their forecasts are likely to be optimistic. However, they may decide not to investigate further and not to correct for possible optimism, because they assume, often for good reason, that optimistic forecasts are more likely to get projects approved than realistic ones. Such behavior has been called the "conspiracy of optimism" (Gardener and Moffat 2008, Hirt 1994, Weston et al. 2007). But this term is an oxymoron, of course. Per definition, optimism is non-deliberate, unintentional, and thus innocent. People who deliberately decide or conspire to be optimistic, are not optimistic; they are practicing strategic misrepresentation and are thus lying.